\RequirePackage[l2tabu, orthodox]{nag}
\documentclass{stsci_report}

\usepackage[pdftex,
pdfauthor={D. Stark},
pdftitle={Improved Identification of Satellite Trails in ACS/WFC Imaging Using a Modified Radon Transform},
pdfsubject={We present a new approach to identify satellite trails (or other linear artifacts) in ACS/WFC imaging data using a modified Radon Transform. We demonstrate that this approach is sensitive to features with mean brightness significantly below the background noise level, and it is resistant to the influence of bright astronomical sources (e.g., stars, galaxies) in most cases. Comparing with a set of satellite trails identified by eye, we find a trail recovery rate of 85 percent and a false detection rate (after removing diffraction spikes that are easily filtered) of 2.5 percent. By performing an analysis using a much larger ACS/WFC data set where false trails are identified by their persistence across multiple images of the same field, we identify the Radon Transform parameter space and image properties where our algorithm is unreliable, and estimate a false detection rate of 10 percent elsewhere.  We apply our method to ACS/WFC data taken between 2002 and 2022 to determine both the frequency of satellite trail contamination in science data and also the typical trail brightness as a function of time. We find the rate of satellite trail contamination has increased by approximately a factor of two in the last two decades, but there is no clear systematic evolution in the typical trail brightness. Our satellite trail identification program is available as part of the {acstools} package.}, 
pdfkeywords={ACS, WFC, CCD, Hubble Space Telescope, HST, Space Telescope Science Institute, STScI, Advanced Camera for Surveys, Wide Field Channel}]{hyperref}

\usepackage{booktabs}
\usepackage{rotating}
\usepackage{natbib}
\usepackage{url}
\usepackage{graphicx}
\usepackage[font=footnotesize,labelfont=bf]{caption}
\usepackage{subcaption}
\usepackage{nameref}
\usepackage{hyperref} 
\usepackage{array} 
\usepackage{color}
\usepackage{amsmath}
\usepackage[export]{adjustbox}[2011/08/13]
\usepackage{threeparttable}
\usepackage{aas_macros}

\copyrighttext{Copyright \copyright \the\year\ The Association of Universities for Research in Astronomy, Inc. All Rights Reserved.}

\title{\textbf{Improved Identification of Satellite Trails in ACS/WFC Imaging Using a Modified Radon Transform}}
\presubtitle{\flushleft{\includegraphics[width=5cm]{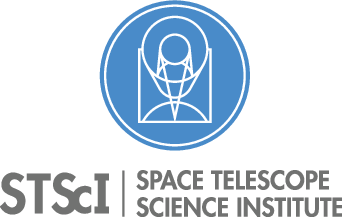}} \\ \hfill Instrument Science Report ACS 2022-08}
\author{David V. Stark, Norman Grogin, Jenna Ryon, Ray Lucas}
\date{December 16, 2022}

\begin{document}

\maketitle

\abstract{We present a new approach to identify satellite trails (or other linear artifacts) in ACS/WFC imaging data using a modified Radon Transform. We demonstrate that this approach is sensitive to features with mean brightness significantly below the background noise level, and it is resistant to the influence of bright astronomical sources (e.g., stars, galaxies) in most cases. Comparing with a set of satellite trails identified by eye, we find a trail recovery rate of 85\% and a false detection rate (after removing diffraction spikes that are easily filtered) of 2.5\%. By performing an analysis using a much larger ACS/WFC data set where false trails are identified by their persistence across multiple images of the same field, we identify the Radon Transform parameter space and image properties where our algorithm is unreliable, and estimate a false detection rate of $\sim10\%$ elsewhere.  We apply our method to ACS/WFC data taken between 2002 and 2022 to determine both the frequency of satellite trail contamination in science data and also the typical trail brightness as a function of time. We find the rate of satellite trail contamination has increased by approximately a factor of two in the last two decades, but there is no clear systematic evolution in the typical trail brightness. Our satellite trail identification program is available as part of the \texttt{acstools} package.}

\section{Introduction} \label{s:intro}
Raw astronomical images are subject to numerous artifacts that must be identified, corrected, or removed before the data can be used for scientific analysis. Among these artifacts are satellite trails, which are frequently seen in HST imaging. With the past and expected future increase in the number of human-made satellites around Earth \citep[e.g.,][]{Lawler2022}, the rate at which HST data are contaminated may similarly increase. In addition to satellite trails, other common artifacts can be linear in nature (e.g., some cosmic rays, certain types of scattered light, charge bleeding, and diffraction spikes), all of which must be detected and masked.

Numerous approaches to identify linear artifacts are present in the literature. In addition to visual inspection \citep{Kruk2022}, a number of automated approaches have been developed including (1) source detection algorithms limited to highly elongated features \citep{Waszcak2017}, (2) machine-learning \citep{Waszcak2017, Paillassa2020, Kruk2022}, and (3) specific image filtering/transformation routines to highlight linear features \citep{Cheselka1999,Keys2010,Virtanen2014,Borncamp2016,Kim2016,Bektesevic2017,Nir2018}. A recent example falling into this latter category, and previously incorporated into the \texttt{acstools} package, was presented by \citet{Borncamp2016}. A Hough transform was used to identify linear edges in an image after it was convolved with an edge-detection filter. More recently, \citet{Nir2018} identified streaks using the Radon Transform \citep[RT, ][]{Radon1917}, an algorithm that calculates the sum of pixels along all possible linear paths crossing an image. 

The RT has numerous advantages over other methods. It acts like a matched-filter for the detection of linear features in an image. For example, the integrated brightness of a satellite trail will be extracted with the highest signal-to-noise ratio ($S/N$) if only the pixels that make up the trail are summed. One can think of the RT as convolving the input image with a linear model (the ``filter") for every possible trail that crosses the image. When the summation occurs along the satellite trail (i.e., when the filter is well-matched to the signal of interest), the RT is maximized. Furthermore, the ability of a linear feature to be detected via the RT does not depend on the local surface brightness of that feature (unlike edge detection-based algorithms) but rather the summed brightness of the feature along a given path. In this sense it is one of the most sensitive algorithms, able to detect features with average surface brightness well below the noise of an image, as long as the feature subtends enough pixels. \citet{Nir2018} make the algorithm even more practical by developing the Fast Radon Transform (FRT) that drastically reduces the calculation time. However, a complication of applying the RT to real data is that an image must first have all other objects (e.g., stars, galaxies) removed. Otherwise, even a single star in an otherwise isolated field will yield a large value in the RT. While one could develop source-finding routines carefully to find astronomical sources while avoiding satellite trails, we explore an alternative approach: instead of calculating the sum along every path crossing an image (the standard RT), calculate the \textit{median} along every path crossing an image (we hereafter refer to this as the Median Radon Transform, or MRT). This approach will be sensitive to continuous linear features but less sensitive to local bright sources. 

In this report, we present a new algorithm to detect satellite trails and other linear features in images using the MRT. We first describe the MRT and its advantages, outline the full program used to identify and mask satellite trails, and then present its sensitivity and performance on real data. Lastly, we use our algorithm to assess how the rate and typical brightness of satellite trails has evolved over the 20 year lifetime of ACS.

\section{The Median Radon Transform} \label{s:observations}
\begin{figure}
\centering
\includegraphics[width=.5\columnwidth,center]{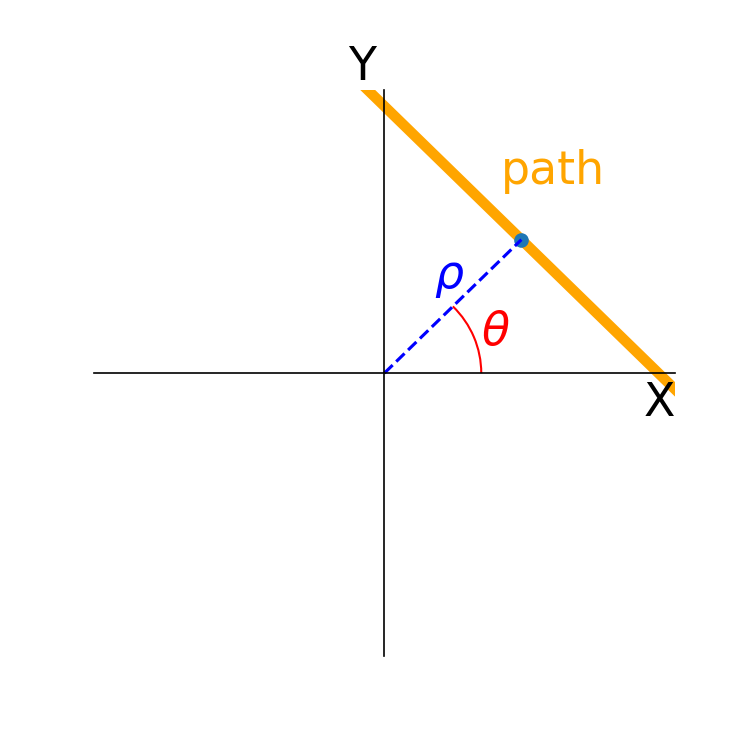}
\caption{The coordinates used to parameterize linear paths over which summations are calculated in the RT. In this example the path of interest (thick orange line) is crossing the $+x,+y$ quadrant and is parameterized by the coordinates $\theta$ and $\rho$, where  $\theta$ represents the angular orientation of the line measured relative to the $x$-axis and $\rho$ represents the distance between the origin and the line. Note that by convention $\theta$ spans $0^{\circ}$ to $180^{\circ}$. The full range of parameter space is filled by having $\rho$ go from $-\infty$ ($-\infty < y < 0$) to $+\infty$ ($0 < y < +\infty$).}
\label{fig:rt_coord}
\end{figure}

\begin{figure}
\includegraphics[width=\columnwidth,center]{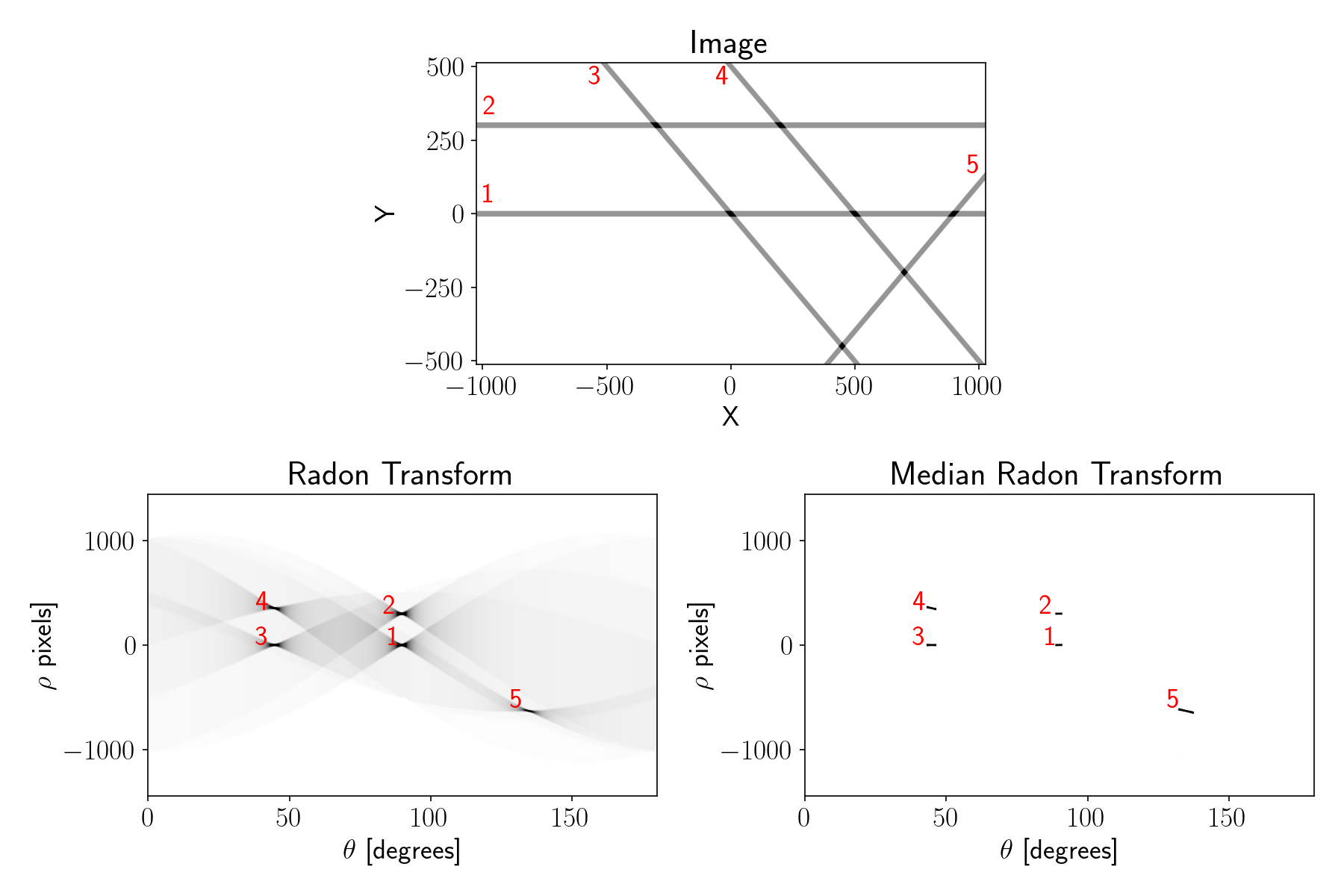}
\caption{(top) A model image with five labeled trails. (bottom-left) Corresponding RT of the image with each trail labeled. (bottom-right) Same as bottom-left but using the MRT. The signals in RT/MRT space are elongated because the model trails are several pixels thick.}
\label{fig:rt_coord_demonstration}
\end{figure}

The standard Radon Transform  operates by summing the values in a 2D function, $f$, over all possible linear paths through that data set
\begin{equation}
R(\theta,\rho) = \int_L{f(x,y)dl}
\end{equation}
The lines crossing an image are parameterized by the coordinates $\theta$, the angle with respect to the positive $x$ axis, and $\rho$, the offset from the origin, typically chosen to lie at the center of an image. The actual path over which the sum is calculated is perpendicular to the line connecting a given ($\theta$,$\rho$) coordinate and the image origin (see Figures~\ref{fig:rt_coord} and \ref{fig:rt_coord_demonstration} for examples).

For the detection of linear features in an otherwise blank image, the Radon Transform is highly advantageous as the sum along the data is maximized along the linear feature and minimized in all other directions. However, when other sources like stars or galaxies are present within the image, the Radon Transform can be elevated by these local bright regions such that the transform is no longer maximized \textit{only} along continuous linear features. While one may avoid this situation by masking these local bright regions, such an approach adds an extra layer of complexity to the analysis and runs the risk of inadvertently masking part or all of a linear feature of interest. We employ an alternative solution: take the median value along each path crossing an image, as opposed to the sum \citep[this solution was also proposed by ][]{Press2006}. We refer to this modified version of the Radon Transform as the Median Radon Transform, or MRT,
\begin{equation}
MRT(\theta,\rho) = \tilde{f}(x,y)_L
\end{equation}
where $\tilde{f}(x,y)_L$ indicates the median of the function $f(x,y)$ over a given path, $L$.

\begin{figure*}
\includegraphics[width=\columnwidth,center]{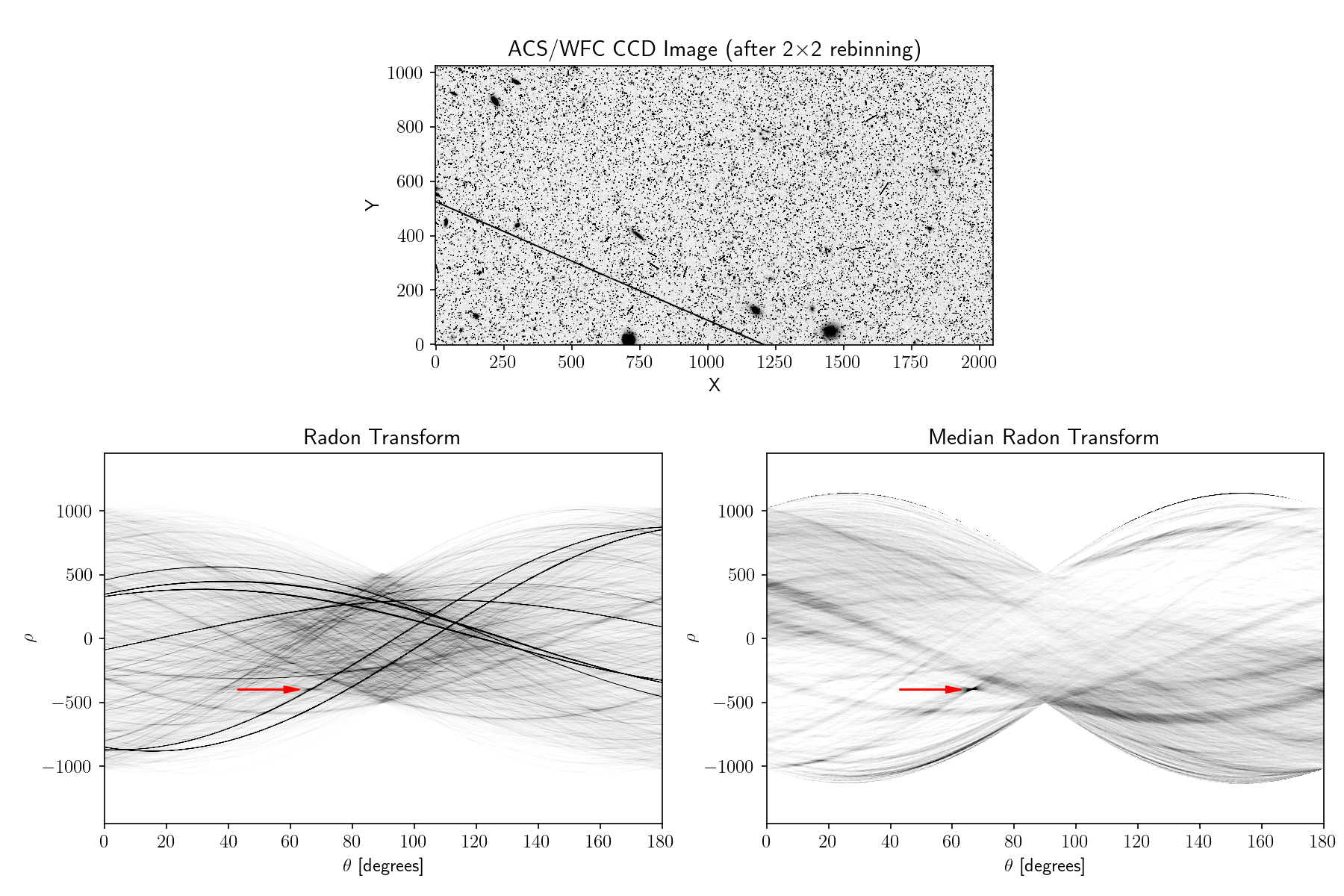}
\caption{A comparison of the standard (bottom-left) and median (bottom-right) Radon Transforms for a single chip of a real ACS/WFC image (rebinned $2\times2$) with a satellite trail (top). Red arrows indicate the location of the signal from the satellite trail in the transformed images. Due to numerous bright astronomical sources, the standard RT is filled with streaks of intensity comparable or brighter than the signal from the satellite trail itself. The MRT significantly reduces the intensity of these additional streaks, making the satellite trail itself the dominant feature in the transformed image.}
\label{fig:rt_image_demonstration}
\end{figure*}

The MRT has the natural advantage of resisting the influence of localized bright sources like stars (as long as the image is background-dominated), while still being sensitive to persistent linear features. The difference in the RT and MRT is apparent in the simple model image presented in Figure~\ref{fig:rt_coord_demonstration} where the RT shows ``flaring" around the signals from the trails, while the MRT signals have sharper edges. Figure~
\ref{fig:rt_image_demonstration} shows the RT and MRT applied to a single chip of a real ACS/WFC image that has a satellite trail, and the difference between the two transforms is even more pronounced. In the MRT, the satellite trail is by far the dominant signal, while in the RT, there are several similar strength signals resulting from the stars and galaxies in the image. 

Unfortunately, the MRT will become less resistant to the influence of astronomical sources in sufficiently crowded fields. This issue will be discussed further in this report (Section~\ref{sec:tests_real_data}).

\section{Trail-finding Algorithm} \label{sec:algorithm}
We have developed a new satellite trail finding algorithm based around the MRT. It is called \texttt{findsat{\_}mrt} and is written in Python. 
Here, we discuss the complete algorithm, including image pre-processing and the culling of false candidate trails extracted from the MRT. This algorithm is applied to ACS/WFC images, but in principle should be applicable to any image data. It was made publicly available as part of v3.6.0 of \texttt{acstools}\footnote{\url{https://acstools.readthedocs.io/en/latest/}}.

\subsection{Image Preprocessing} \label{sec:preprocessing}
Throughout this report, we run \texttt{findsat{\_}mrt} on ACS/WFC \texttt{flc} images created by the \texttt{calacs} pipeline\footnote{The individual frames have typically been incorporated into drizzled data products, so their \texttt{DQ} arrays have additional flags beyond what is set by \texttt{calacs}.} and each of the two ACS/WFC CCDs is analyzed separately. Bad pixels are identified using the \texttt{DQ} arrays and set to \texttt{NaN} to reject them from further analysis. We do not remove pixels flagged as cosmic rays (\texttt{DQ} flags \texttt{4096} and \texttt{8192}), as these flags have been found to partially flag satellite trails, hindering our ability to detect them. Using only good pixels, a median background is subtracted from each image. 

We also apply $2\times2$ rebinning on all images. We generally recommend this approach when analyzing ACS/WFC images because doing so decreases the calculation time of the MRT on WFC data by an order of magnitude. We do not see any evidence that our rebinning scheme affects our ability to detect satellite trails, which are typically several pixels wide.

\subsection{MRT Calculation and Source Identification}
When the MRT is being calculated on a given image, any masked image pixels are ignored when the medians are calculated along each linear path. A length array, $L$, which specifies the number of valid pixels used to calculate the MRT at each position, is simultaneously recorded. Any coordinates with $L<25$ are flagged due to the median being poorly constrained with so few points.

The MRT calculation employed here is based on the \texttt{skimage.transform.radon} routine \citep{scikit-image} but with the summation replaced by a median. The MRT is computationally intensive, and although there are proposed methods of speeding up the algorithm \citep[see e.g., ][]{Press2006,Nir2018}, for now we have taken the approach of using the \texttt{multiprocessing} package. Using 24 3.10 GHz cores, calculating the MRT on a $2048\times1024$ image (the size of a single ACS/WFC CCD after $2\times 2$ binning) with an angular sampling of $0.5^{\circ}$ takes $\sim$10 seconds (this angular sampling appears sufficient to identify trails in our data set).

We calculate an approximate MRT uncertainty map as
\begin{equation} \label{eq:sigma_mrt}
\sigma_{\rm MRT}(\theta,\rho)=\frac{1.25\sigma_{\rm im}}{\sqrt{L(\theta,\rho)}}
\end{equation}
where $\sigma_{\rm im}$ is an outlier-resistant image standard deviation. Equation~\ref{eq:sigma_mrt} represents the standard error on a median assuming a Gaussian distribution. While a more robust estimate of the uncertainty at each position may be accomplished using a method like bootstrap resampling, such an approach would be computationally prohibitive at this point.

Regions in the MRT with $S/N$ above a certain threshold (5 by default) are identified within the MRT using a source finder (the \texttt{photutils.Detection.StarFinder} Python module) and kernels derived from MRTs of mock satellite trails. As trails can have different widths, we use three kernels with widths of 15, 7, and 3 pixels. 

\subsection{Refining the Source List}
Source detection in the MRT may often yield several candidate trails, but even with its resistance to non-continuous features, not all candidates are robust or relevant to our search for linear artifacts. Additional steps are therefore taken to cull the list of spurious detections. First, we extract a cutout around each candidate trail and rotate it so the trail runs in the horizontal direction, along the rows of the rotated cutout. We use bi-linear interpolation to resample the data after rotation. The median is then taken along each row to create a one-dimensional cross-section of the trail. Background variations are removed by performing a third order polynomial fit to the data, avoiding the central pixels where we expect the trail to be found\footnote{We set the width of the avoided central region to be equal to the maximum trail width allowed. In our case, that is 75 pixels.}. A Gaussian function is then fit to the data to refine the estimate of the trail width and position. The trail width is calculated as $w=2\times(3\sigma)$ where $\sigma$ is the derived standard deviation from the Gaussian fit. The peak intensity is calculated from the data within $+/-3\sigma$ of the Gaussian centroid. The background root mean squared (rms) noise is calculated by taking a standard deviation of all data outside $+/-3\sigma$ of the Gaussian centroid. We use the peak brightness and baseline rms noise to then re-estimate the $S/N$ of the trail. 

A trail candidate is kept if (1) its re-estimated $S/N$ is above a set threshold (5 by default), and (2) it is within a pre-defined maximum width. The maximum width, which is 75 pixels for our analysis of ACS/WFC data\footnote{This value was chosen from trial and error on images with satellite trails and can be adjusted.}, is effective at removing detections of semi-continuous linear structures formed by bright stars and larger galaxies (although we will see in Section~\ref{sec:tests_real_data} that this issue is not completely avoided), as well as large scale background variations. All of the parameters described above can be modified by the user. 

As satellite trails are expected to cross from one side of an image to the other, we also measure what we call the ``persistence" of a trail, that is, how continuously it can be detected across an image. For each candidate trail that is detected in the MRT and passes the filtering step described above, we again extract a cutout which is then broken into $N$ sections along the length of the trail. The number of sections is determined by estimating how many sections a uniform trail can be broken into and still have signal-to-noise per section $S/N_i$ $ \ge 3$, assuming $S/N_i$ can be estimated as:
\begin{equation}
S/N_i = \frac{S/N_{\rm total}}{\sqrt{N}}
\end{equation}
leading to 
\begin{equation}
N=\left(\frac{S/N_{\rm total}}{S/N_i}\right)^2 = \left(\frac{S/N_{\rm total}}{3}\right)^2
\end{equation}
In practice we round $N$ down to the lowest integer for each trail. We also require the sections to be at least 100 pixels long.

The data in each section is collapsed into a 1D cross-section and fit with a Gaussian function as described above. If a fit has $S/N > 3$ and width within the maximum allowed value, it contributes $+1$ to the persistence score. The fits to each trail subsection are done serially, with the best-fit of the previous section being used as the starting guess of the next, unless the prior fit failed or was significantly different from the fit before that. The final persistence is the total persistence score divided by the number of sections, $N$. We generally reject anything with a persistence score $<0.5$. 

\section{Performance}
We now present tests of the performance of the trail-finding algorithm laid out in Section~\ref{sec:algorithm}. We first discuss its theoretical sensitivity and performance on idealized simulations. We then compare the trails we identify with those identified by-eye from the Frontier Fields program \citep{Lotz2017}. Finally, we examine the distribution of trails recovered in several years-worth of ACS/WFC data to identify regions of parameter space where the MRT-based algorithm returns reliable results.

\subsection{Sensitivity}
The MRT relies on measurements of the median along different paths across an image. The standard error on the median flux is approximated using Equation~\ref{eq:sigma_mrt}. Due to its dependence on the number of pixels, the sensitivity of the MRT to satellite trails will vary across the image, being most sensitive to trails that cut through the center of the image. If we denote $\sigma_{\rm im}$ as the background noise, then the median trail flux detected at some $S/N$ relative to the background noise, $s=\tilde{f}/\sigma_{\rm im}$, is given as
\begin{equation} \label{eq:sensitivity}
\frac{\tilde{f}}{\sigma_{\rm im}}=\frac{1.25 (S/N)}{\sqrt{L}}
\end{equation}
where $S/N=\tilde{f}/\sigma_{\tilde{f}}$, and $\sigma_{\tilde{f}}$ is the uncertainty on the median. For the longest pathlength through a single ACS chip ($L=4580$ pixels), we expect the MRT to yield $S/N=5$ detections of trails with median brightness $\sim 0.1$ times the background noise. Near the corners of the chip, for example, with $L=500$ pixels, the sensitivity at $S/N=5$ should be $\sim0.3$ times the background noise. Therefore, over the majority of the WFC CCD, the MRT should be sensitive to trails well-below the noise level.

To test the theoretical sensitivity, taking into consideration various detector imperfections, we ran a set of ACS/WFC simulations on observations of empty sky (with a Poisson-variable uniform sky level) and artificially generated satellite trails. The trails are modeled as being 5 pixels wide (typical for trails identified by eye) and are convolved with a Gaussian with full width at half maximum of 2.5 pixels to approximate the PSF. Once the sky model is generated, the simulation adds in various CCD effects: read noise, imperfect charge-transfer, non-uniform flatfield response, dark current, and bias level. These images are then passed through the ACS pipeline to create mock \texttt{flc} images\footnote{We closely follow the examples at \url{https://github.com/spacetelescope/acs-notebook/blob/master/acs_cte_forward_model/acs_cte_forward_model_example.ipynb} for this step.}. We use an integration time of 350 seconds with a typical F606W filter sky illumination of 125.59 electrons per pixel per kilosecond \citep{Anand2022}.

We perform two simulations, one with a trail crossing diagonally along the longest WFC pathlength for a single chip, and one crossing a corner with a length of 500 pixels. We repeat the simulations varying the trail peak brightness from 0.05 to 0.5 times the background noise level in steps of 0.01, each time running the MRT trail detection algorithm. We find $S/N \ge 5$ detections of the trails in the MRT down to 0.13 and 0.37 times the noise level for the simulations with the long and short trails, respectively. For comparison, the \texttt{detsat} routine previously incorporated into \texttt{acstools} detects the long trail consistently down to a brightness of 1.2 times the background level using the default parameters. The short trail is not consistently detected by \texttt{detsat} even up to 3 times the background level, but becomes fully undetectable below 1.6 times the background noise level. 

The tests presented here provide an idealized understanding of the sensitivity of the MRT, but real images with astronomical sources (e.g., stars, galaxies) can complicate detection. Next, we examine how \texttt{findsat{\_}mrt} performs on real data, and what additional steps are necessary to isolate real satellite trail signals.

\subsection{Tests on Real Data} \label{sec:tests_real_data}

When assessing the reliability of our satellite trail finder on real astronomical images, we try to answer the following questions: (1) What fraction of satellite trails are properly recovered (or equivalently, what fraction of true satellite trails are missed), (2) what fraction of identified satellite trails are false, i.e., from something other than a satellite, and (3) what causes false trails, and do they occupy regions of parameter space that can be easily rejected?

We assess the reliability of our code using two approaches. The first compares the recovered trails to those visually identified within the Frontier Fields. A second approach uses a larger data set and looks for abnormal excesses of identified trails in parameter space, particularly for duplicate trails (i.e., those that occur in the same place over multiple observations of the same field) that are expected to be ``false" satellite trail detections. 

\subsubsection{Comparison with By-Eye Classifications}
\begin{figure}
\includegraphics[width=1.1\columnwidth,center]{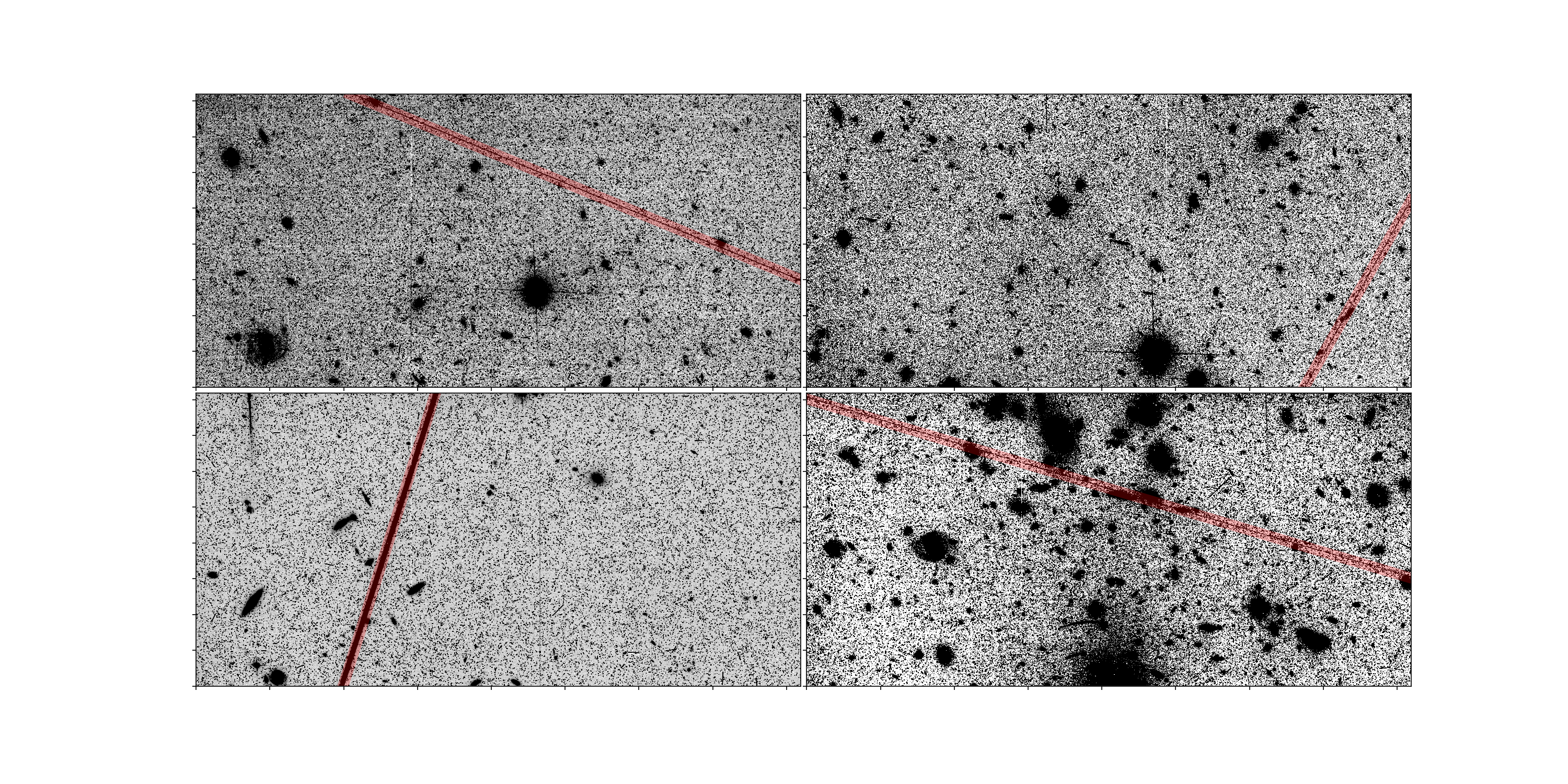}
\caption{Examples of trails identified by both \texttt{findsat{\_}mrt} and previous by-eye inspection. The red bands highlight the trails.}
\label{fig:recovered_trails}
\end{figure}

\begin{figure}
\includegraphics[width=1.1\columnwidth,center]{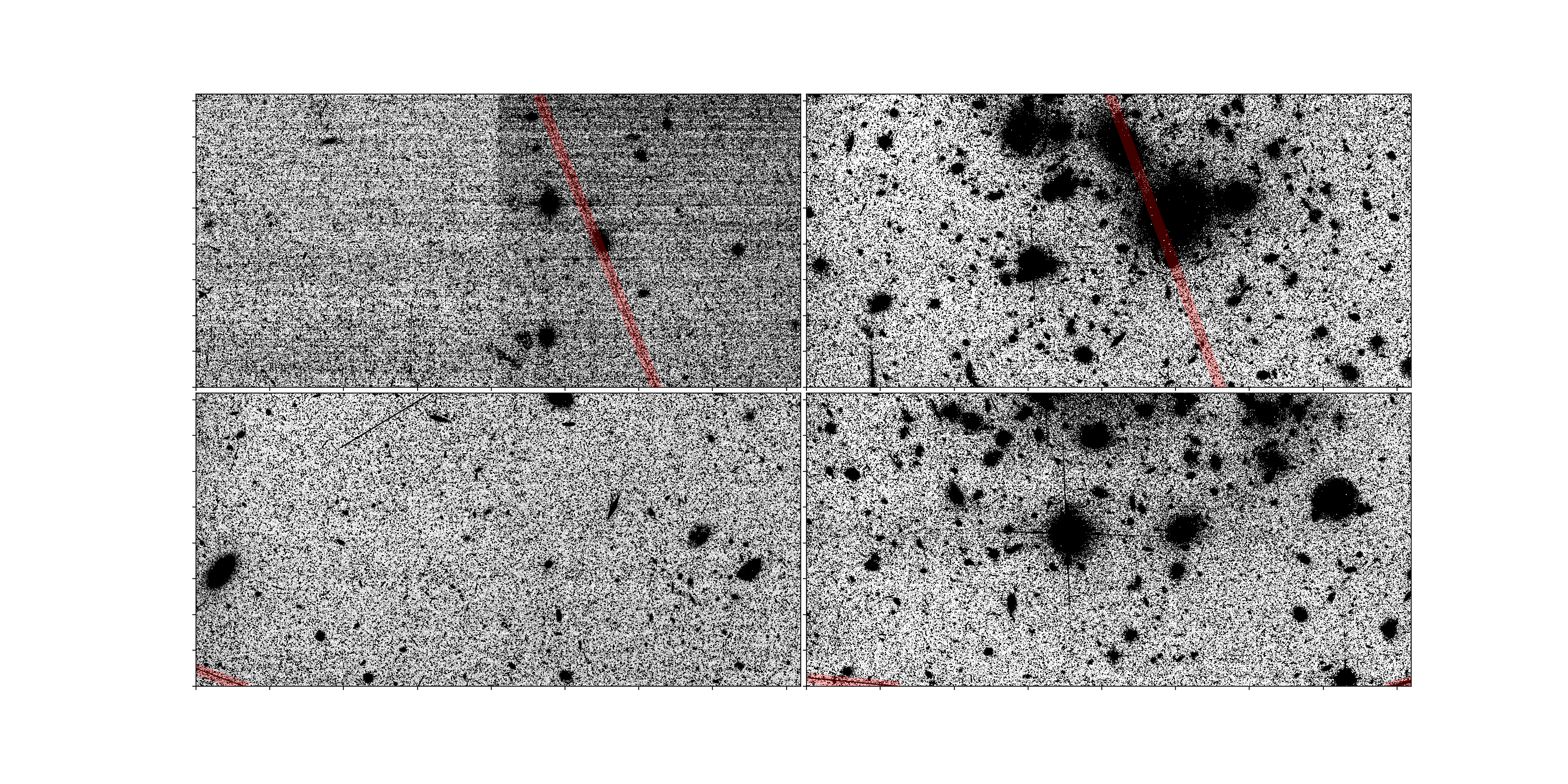}
\caption{Examples of trails found via by-eye inspection but missed by \texttt{findsat{\_}mrt}. The red bands highlight the by-eye trails. (top-left) The trail was identified but later rejected due to having persistence $<0.5$. (top-right) The trail was identified but rejected due to being too wide, likely due to the alignment with the bright galaxies in the frame. (bottom panels) Trails passing through the corners of the images. Roughly half of missed trails fall close to the corners.}
\label{fig:missed_trails}
\end{figure}

\begin{figure}
\includegraphics[width=1.1\columnwidth,center]{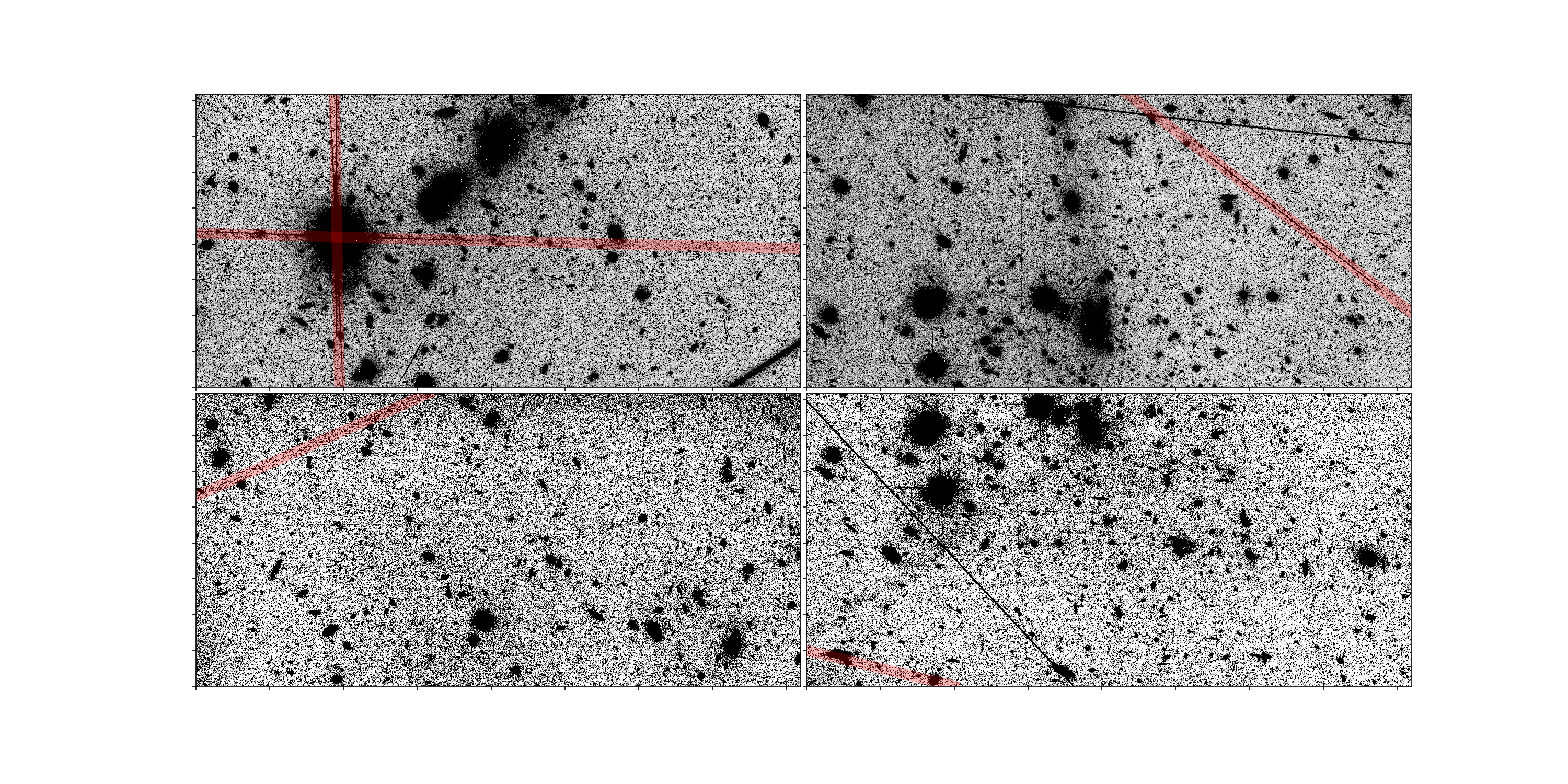}
\caption{Examples of trails found by \texttt{findsat{\_}mrt} that were not present in the by-eye masks created for the Frontier Fields images. The red bands highlight the trails. (top-left) A case where diffraction spikes were flagged. (top-right) A long cosmic ray was identified. (bottom-left) A real trail that was missed by the original Frontier Fields by-eye image inspection. (bottom-right) A false trail formed by the alignment of astronomical sources.}
\label{fig:false_trails}
\end{figure}

We ran our algorithm on 358 Frontier Fields images with corresponding by-eye satellite masks. Our results are summarized in Table~\ref{tbl:ff}. We show examples of recovered, missed, and newly identified trails in Figures~\ref{fig:recovered_trails}, \ref{fig:missed_trails}, and \ref{fig:false_trails}.

\begin{table} 
\centering
\begin{threeparttable}
\caption{Results of running \texttt{findsat{\_}mrt} on Frontier Fields Images}
\begin{tabular}{ || l r ||}
\hline
By-eye trails recovered$^a$ & 255 (267,261)\\ 
By-eye trails missed$^a$ & 47 (35,41)\\   
\hline
Total New Trails Found & 46 \\
\hspace{1cm}$\rightarrow{}$Diffraction spikes & 37 \\
\hspace{1cm}$\rightarrow{}$False trails & 7 \\
\hspace{1cm}$\rightarrow{}$Real trails & 2 \\
\hline
\end{tabular}
\begin{tablenotes}
\footnotesize
\item $^a$ The numbers given refer to the number of trails after the catalog was trimmed of low $S/N$, wide, and/or low-persistence trails. The values in the parentheses provide the numbers (1) after detection in MRT only and (2) after filtering low $S/N$ or wide trails, but not filtering on persistence.
\end{tablenotes}
\label{tbl:ff}
\end{threeparttable}
\end{table}

We successfully recover 255 out of 302 trails (Figure~\ref{fig:recovered_trails}), with 47 being missed (Figure~\ref{fig:missed_trails}). Of the missed trails, six were identified in the first stage of the algorithm that looks for $S/N$ peaks directly in the MRT, but they were rejected in subsequent filtering based on a reassessment of the $S/N$ and trail width. An additional six trails that passed the first round of filtering were subsequently rejected based on their measured persistence being $<0.5$. Of the remaining 35 missed trails, approximately half of them were very short (typically a few hundred pixels long) and found in the corners of the images where the MRT is least sensitive. The overall recovery rate is 85\% under the strictest definition (considering only trails that passed all filtering stages) or 88\% under a more relaxed definition (anything found by the first stage of the MRT). It is worth noting that some of the features in the ``by-eye" masks are diffraction spikes, not satellite trails. We find diffraction spikes are not consistently masked in the Frontier Fields images. Our algorithm sometimes removes diffraction spikes with the persistence analysis if their brightness sufficiently dims, away from their origin. This issue accounts for some of the ``missed" trails.

Our algorithm also finds 46 additional trails not present in the by-eye masks (see Figure~\ref{fig:false_trails}). The vast majority of these trails (37/46) can be easily attributed to diffraction spikes, i.e., they fall at angles around $0/180^{\circ}$ and $90^{\circ}$ expected for diffraction spikes. The remaining images were inspected by eye to determine the cause of the additional trails. In two cases, weak trails were found that were missed in the by-eye inspection. In another case, the trail was actually a long cosmic ray. The remaining cases are mostly found in corners of images and appear to be alignments of astronomical sources that make pseudo-continuous linear ``trails" that \texttt{findsat{\_}mrt} cannot distinguish from real satellite trails. Thus, if we ignore diffraction spikes (which are relatively easily avoided), there are seven false trails found implying a false detection rate of 2.5\%.

The analysis here suggests the MRT trail finder recovers the majority of trails identified by eye, but struggles with completeness and false detections in the corners of images where trails are short. It cannot always distinguish other linear image artifacts like some cosmic rays and diffraction spikes, although it will still be useful to flag such features in general. We investigate these trends further with more data in the following section. 

\subsubsection{Performance on Multi-Exposure Fields}

We next analyze results from applying \texttt{findsat{\_}mrt} over several years of ACS operations. This approach expands the range of image properties being tested, enabling a more robust assessment of our algorithm's reliability and limitations. The caveat of this approach is that we do not have complete by-eye classifications of satellite trails, so we cannot explicitly determine which trails are real versus spurious. To address this limitation, we work under the assumption that no true satellite should cross the exact same location on an image more than once when the telescope is pointing at the same area on the sky. 

Any trails that appear in the same location in multiple images of the exact same region on sky are referred to as  ``duplicates" and assumed to \textit{not} be satellite trails, which should be stochastic (referred to as ``unique"). In order to be able to flag unique versus duplicate trails, we limit our analysis to fields with at least two exposures at the same coordinates (aperture Right Ascension and Declination changing by no more than 5") and telescope orientation (aperture position-angle changing no more than 5$^{\circ}$). Any trails from images pointing at the same area and whose $\theta$ and $\rho$ measurements are within $5^{\circ}$ and 75 pixels of each other are flagged as ``duplicate trails", and presumed to be false satellite trails (although they may represent other real artifacts). By plotting the distributions of unique and duplicate trail properties (e.g., position, length, host image characteristics), and comparing to theoretical expectations, we can identify regions of parameter space with trail excesses or deficits, respectively, indicative of high false-positive rates or low completeness.

Our analysis incorporates ACS/WFC imaging data from 2-3 year intervals spanning 2002 to 2022. We include all full-frame external-target images over all filters with exposure time $>0$ seconds.

\begin{figure}
\includegraphics[width=\columnwidth,center]{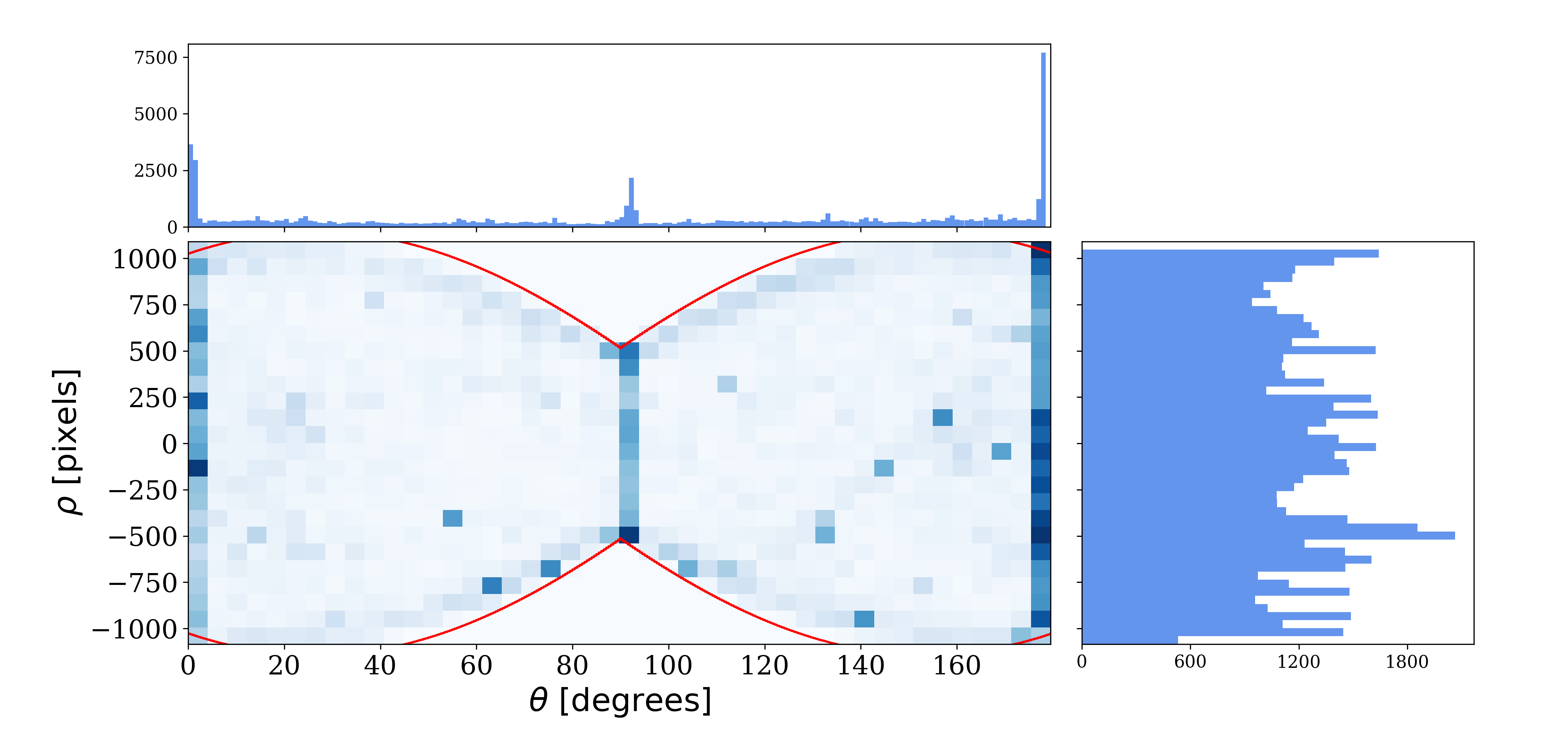}
\caption{Distribution of $\theta$,$\rho$ for trails identified in ACS/WFC data. The red lines denote the edge of valid parameter space. The peaks around $\theta$=0, 90, and 180 degrees correspond to diffraction spikes.}
\label{fig:rho_theta_distribution}
\end{figure}

We first plot the distributions of trail angle ($\theta$) and offset ($\rho$) for all detected trails (see Figure~\ref{fig:rho_theta_distribution}). There are clear excesses of trails at angles of $\sim0^{\circ}$, $\sim90^{\circ}$, and $\sim 180^{\circ}$. As with our analysis of the Frontier Fields, visual inspection shows that these trails arise mostly from diffraction spikes, which can often cross the whole CCD and, if bright enough, can pass the persistence test we apply to all detected trails. Given the overwhelming dominance of diffraction spikes at these angles, we conclude that any trails identified around these angles are highly unlikely to be satellite trails. We flag and remove them from subsequent analysis. 

We find no clear excess in the density of trails with respect to trail offset, $\rho$, alone. However, in $\theta,\rho$ space, there are random localized excesses in identified trails, and there is an excess along the maximum possible absolute offsets at all angles (around the outer-edge of the 2D distribution in Radon space). These coordinates correspond to trails cutting through the corners of images.

\begin{figure}
\includegraphics[width=0.5\columnwidth,center]{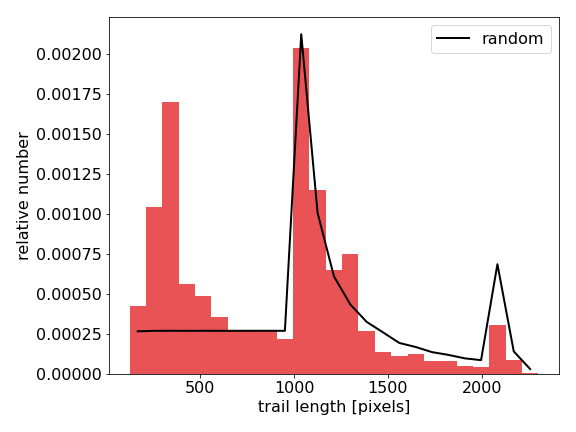}
\caption{Distribution of lengths for all trails found in ACS/WFC data. The black line shows the expectation from a purely random distribution of trails in image space.}
\label{fig:length_distribution}
\end{figure}

\begin{figure*}
\includegraphics[width=\columnwidth,center]{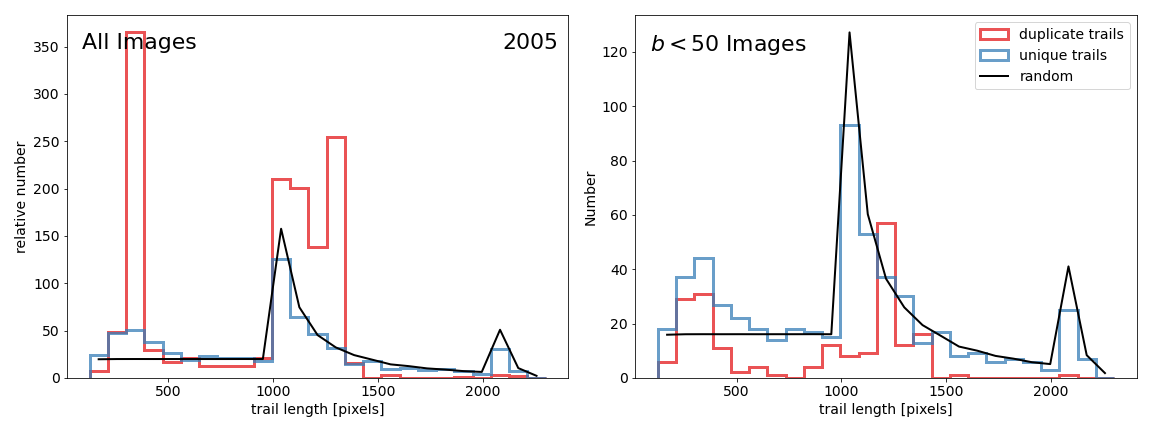}
\includegraphics[width=\columnwidth,center]{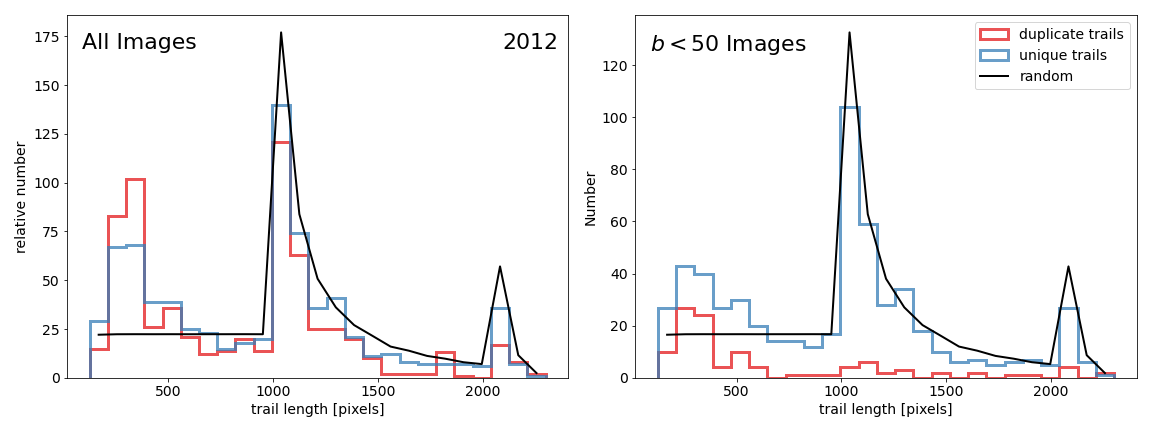}
\caption{Distribution of lengths for unique (presumed real) and duplicate (presumed false) satellite trails in 2005 and 2012 ACS/WFC data. For comparison we overplot the expected distribution from trails with random offsets and orientations. The left panel shows the distribution of trails found in all images, while the right panel shows only the distribution of trails from $b<50$ images. Generally, the $b<50$ selection rejects many of the duplicate satellite trails especially at large lengths. The remaining excess of trails at $L\sim1200$ pixels in the 2005 data is an exception (see text).}
\label{fig:trail_lengths_years}
\end{figure*}

Trails cutting through image corners correspond to those with shorter lengths, so we next examine how the number of trails is distributed as a function of trail length (see Figure~\ref{fig:length_distribution}). For comparison, we also plot the expected distribution of trail lengths assuming completely randomized satellite trail positions and orientations. This distribution was taken straight from the $L$ array returned by \texttt{findsat{\_}mrt} that provides the trail length at every sampled ($\theta,\rho$) coordinate in the MRT. The comparison of our distribution and the expected distribution shows that there is a strong excess of detected trails at lengths $<500$ pixels.  To clarify the cause of the excess, we separate the unique and duplicate trails in Figure~\ref{fig:trail_lengths_years}. From this figure, it is clear that an excess number of trails at short lengths is largely driven by duplicate trails, but we also see large numbers of duplicate trails with longer lengths. As hinted by our analysis of trails found in Frontier Fields imaging, further visual inspection suggests that a majority of duplicate trails are caused by chance alignments of stars and galaxies that form pseudo-continuous trails across images. While the brightness distribution along such ``trails" is not continuous, if the separation between astronomical sources is small enough, our code cannot differentiate between them and a true continuous satellite trail. 

False trails formed by aligned astronomical sources should be a more pronounced issue in denser fields (e.g. star clusters) where linear ``paths" made of stars/galaxies can be more easily found. To confirm this expectation, we measure a skewedness-like parameter in the pixel brightness distribution, $b$, of all images, here defined as 
\begin{equation}
b=\frac{q_4-q_2}{q_2-q_1}
\end{equation}
where $q_N$ represents the value of the Nth quartile of pixel values in the image\footnote{$q_4$ represents the 99th percentile of the data, not the 100th percentile, so as to avoid extreme outlier pixels}.  This metric is designed to capture the fraction of an image dominated by bright sources, as opposed to background. 

\begin{figure}
\includegraphics[width=\columnwidth,center]{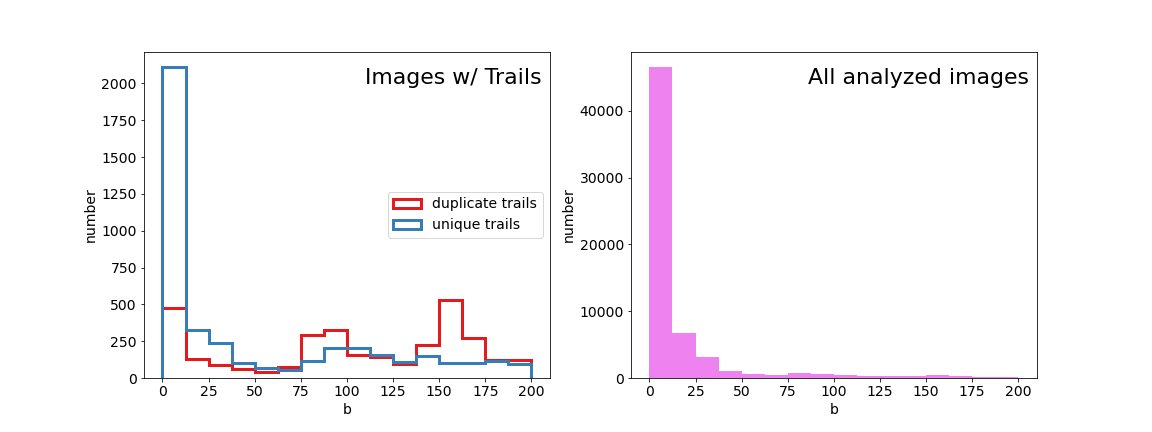}
\caption{(left) Distribution of image brightness distribution skewedness parameter, $b$ for host images of unique and duplicate trails. Duplicate trails are dominant above $b\sim50$. (right) Distribution of skewedness parameter for all analyzed images. The majority of images have $b<50$, implying the majority of false satellite trails result from the smaller fraction of images of very high density fields. }
\label{fig:trail_skewedness}
\end{figure}

Figure~\ref{fig:trail_skewedness} (left panel) shows the distribution of $b$ for unique and duplicate trails. Above $b\sim50$ the rate of duplicate trails is much higher than at $b<50$. This finding is consistent with our expectations and motivates our general practice of avoiding using any trails identified in $b>50$ images. The vast majority of analyzed images have $b<50$ (right panel of Figure~\ref{fig:trail_skewedness}), so such a restriction will not affect the majority of ACS/WFC data.

The right panels of Figure~\ref{fig:trail_lengths_years} show the resulting distribution of unique and duplicate trail lengths after rejecting all images with $b>50$. The fraction of duplicate trails are significantly reduced, but two potential issues remain. First, there is still an excess, albeit smaller than before, of false trails at short lengths. This excess is seen in both the duplicate {\it and} unique trails. The excess of duplicate trails appears to still be due to aligned astronomical sources. These occurrences can be understood by considering that it takes fewer sources to generate a pseudo-random trail in image corners, increasing the likelihood that this will occur, even in lower density fields. The excess in unique trails needs further investigation, but in multiple exposures of the same field, trails are inconsistently found in corners. This behavior may be related to varying noise levels in images of the same field with different exposure times.

The second potential lingering issue is seen in the top-right panel of Figure~\ref{fig:trail_lengths_years} where we show a remaining excess of duplicate trails with lengths around $\sim1200$ pixels in data from 2005. By-eye investigation showed that this excess is almost entirely due to a ``glint" caused by a bright source lying near/within the ACS/WFC chip gap (see \href{https://hst-docs.stsci.edu/acsdhb/files/60243156/60243160/1/1567011424969/acs_Ch46.4.jpg}{Figure 4.25 of the ACS DHB}). The glint occurs in several repeat UDF-05 follow-up observations of the Hubble Ultra Deep Field \citep{Oesch2007}. While this glint is still an artifact worth identifying and flagging, it does mean that an assessment of genuine satellite trails is still subject to contamination after restricting to trails with $b<50$ and $L>500$ pixels. However, the behavior seen in the 2005 data appears to be an exception due to extremely deep repeat-imaging of a particular field. 

Not every field observed with HST will have multiple exposures. In such cases, our analysis shows that trails identified at lengths $<$ 500 pixels and $b>50$ should be used with caution, as they are highly likely to be spurious. For trails with lengths $>500$ and $b<50$, we estimate the false detection rate for satellite trails using the ratio of duplicate trails to total trails. We find this ratio to be 13\%. If we ignore the 2005 data that shows the abnormally high numbers of observations suffering from scattered light, the false rate drops to 10\%. However, as noted earlier, some of the features identified as trails that are not true satellite trails may still be features worth masking.

\subsubsection{Recovering Additional Trails that Cross Multiple Chips}
As discussed in Section~\ref{sec:preprocessing}, \texttt{findsat{\_}mrt} was run on each WFC chip separately. However, the proximity of the two chips means that approximately half of all satellite trails should cross both of them. Identifying such multi-chip trails has a few key uses. First, identifying a trail that crosses both chips provides useful confirmation that a trail detected in an individual  chip is robust, as it is highly unlikely to have two false detections that are exactly aligned in both chips. Connecting trails across multiple chips provides a means of recovering those that have low $S/N$ or lie in an unreliable region of parameter space (e.g., L$<$500 pixels). Second, accounting for multi-chip trails is a necessary step when determining the rate of satellite trail contamination. When both WFC chips are analyzed separately, a single trail crossing both chips will be incorrectly counted as two independent trails. 

To identify trails crossing both chips, we iterate through each trail in WFC1, identifying the point where each trail reaches the chip gap, then interpolate its starting position on WFC2. Our interpolation ignores that there is significant distortion in the WFC detector (see \href{https://hst-docs.stsci.edu/acsihb/chapter-10-imaging-reference-material/10-4-geometric-distortion-in-acs#id-10.4GeometricDistortioninACS-10.4.1WFC}{Section 10.4 of the ACS IHB}). Therefore, we search for trails in WFC2 starting from our predicted position $\pm$ 100 pixels to account for this uncertainty. Despite the distortion, a single trail should have the same angle, $\theta$, on  both chips. Therefore, we also require any matched trails to have the same angle within five degrees. This ``multi-chip" technique only requires that a trail be detected in the MRT, not that it passes the additional filtering stages based on width and persistence.

 In our analysis of ACS/WFC data, we find a total of 982 trails using the multi-chip technique that do not have angles corresponding to diffraction spikes (these equate to 491 unique trails given that a trail that crosses both chips will be counted twice). Of these multi-chip trails, 81 are flagged as duplicates, implying a false-positive rate from the multi-trail identification of $\sim$8\%. 883 ($\sim$90\%) of the multi-chip trails are also identified by our standard technique based on $S/N$, trail width, and persistence, although this number would drop to 785 ($\sim80\%$) using the more strict application of the standard method where any trail with $L<500$ pixels or $b>50$ is ignored due the high rate of false positives. Therefore, identifying trails in multiple chips does indeed present a way of recovering a significant number of trails that would otherwise be missed. 

 At this stage, we do not use the multi-chip approach to {\it reject} trails that are found in only one chip but should, based on their trajectory, be detected in both. The reason for this choice is that to connect any trail across both chips it has to be detectable in each individual chip, but our sensitivity to trails is highly dependent on trail length (Eq.~\ref{eq:sensitivity}). For example, a trail's length may be short in one chip (for example, if it crosses a corner near the chip gap) and longer in the other chip. The shorter component may not be detectable due to the fewer pixels available, but the trail is nonetheless still real.  

In the future, we plan to run \texttt{findsat{\_}mrt} on distortion-corrected images that contain both chips in the same image extension (\texttt{drc} files). This approach will remove the need to match trails that cross both chips, and will increase the overall sensitivity of the algorithm by increasing the potential length of each trail. From a practical point of view, this approach has negligible impact on the computation time of the MRT.

\section{Summary and Recommendations for Use}
We summarize our analysis on the performance of the MRT-based satellite trail finding algorithm, \texttt{findsat{\_}mrt} as follows:
\begin{itemize}
\item In principle, it is sensitive to trails with average brightness as low as $\sim0.1$ times the background noise, although the sensitivity is dependent on the trail length and will decrease towards the corners of images. Our algorithm is approximately an order of magnitude more sensitive than the current algorithm offered by the ACS team to identify ACS/WFC satellite trails.
\item It recovers $\sim85\%$ of trails identified through visual inspection, with a false detection rate (after removing angles corresponding to diffraction spikes) of $\sim$2.5\%. This test was performed on the Frontier Fields data set so the rates may change in images with different properties.
\item It tends to miss trails in image corners where sensitivity is lowest, but also tends to find false trails in corners of images, particularly where trail lengths are $<500$ pixels and few astronomical sources are needed to create pseudo-continuous trails across the image.
\item It cannot differentiate certain other linear artifacts from satellite trails. Bright/long diffraction spikes, cosmic ray hits, or glints can be detected. While not satellite trails, such features may still be useful to mask.
\item Images with high densities of astronomical sources can lead to more false trails due to our algorithm connecting astronomical sources into a single path, even at large trail lengths. Images with $b<50$ yield mostly reliable trails at $L>500$ pixels. 
\end{itemize}
Based on this performance, we recommend the following for users:
\begin{itemize}
\item Trails identified in fields with high source density $b>50$ should be inspected. These are more frequently the result of chance arrangements of stars and galaxies creating pseudo-continuous trails across images.
\item For similar reasons, any trails at lengths $<500$ pixels, regardless of image source density, should be inspected. 
\item If the goal is explicitly to find satellite trails, the range of angles corresponding to diffraction spikes should be ignored. 
\item Multiple observations of a field provide a valuable check to determine whether a detected feature is truly a satellite trail. In the absence of duplicate images, certain artifacts (scattered light/glint) may be difficult to differentiate from satellite trails.
\item If analyzing data from multi-chip detectors (like ACS/WFC) identifying trails that cross multiple chips provides another means of robust detection. However, the process of connecting such trails can be circumvented by analyzing all chips simultaneously (e.g., using \texttt{drc} files for WFC).
\end{itemize}

\section{Results: The Evolution of Satellite Trail Frequency and Brightness in ACS/WFC data}

Now that we have assessed regions of parameter space where \texttt{findsat{\_}mrt} provides a clean sample of satellite trails, we assess how the trail frequency in ACS/WFC data has evolved over time. To generate as clean a sample of satellite trails as possible, we apply the following selections to the trail catalog:
\begin{itemize}
\item We require $S/N > 5$, maximum trail width $<$ 75 pixels, and trail persistence $>$ 0.5
\item We remove any trails found in angles populated by diffraction spikes
\item We limit the analysis to low density fields with $b<50$
\item We reject trails with lengths $>500$ pixels
\item We make exceptions to the above selections if a trail crosses both chips. 
\item We require at least two exposures per field, and reject any trails flagged as duplicates\footnote{ACS observations typically have multiple observations per field for cosmic ray/bad pixel removal, so this latter requirement rejects very little data.}
\end{itemize}

\begin{figure}
\centering
\includegraphics[width=\columnwidth,center]{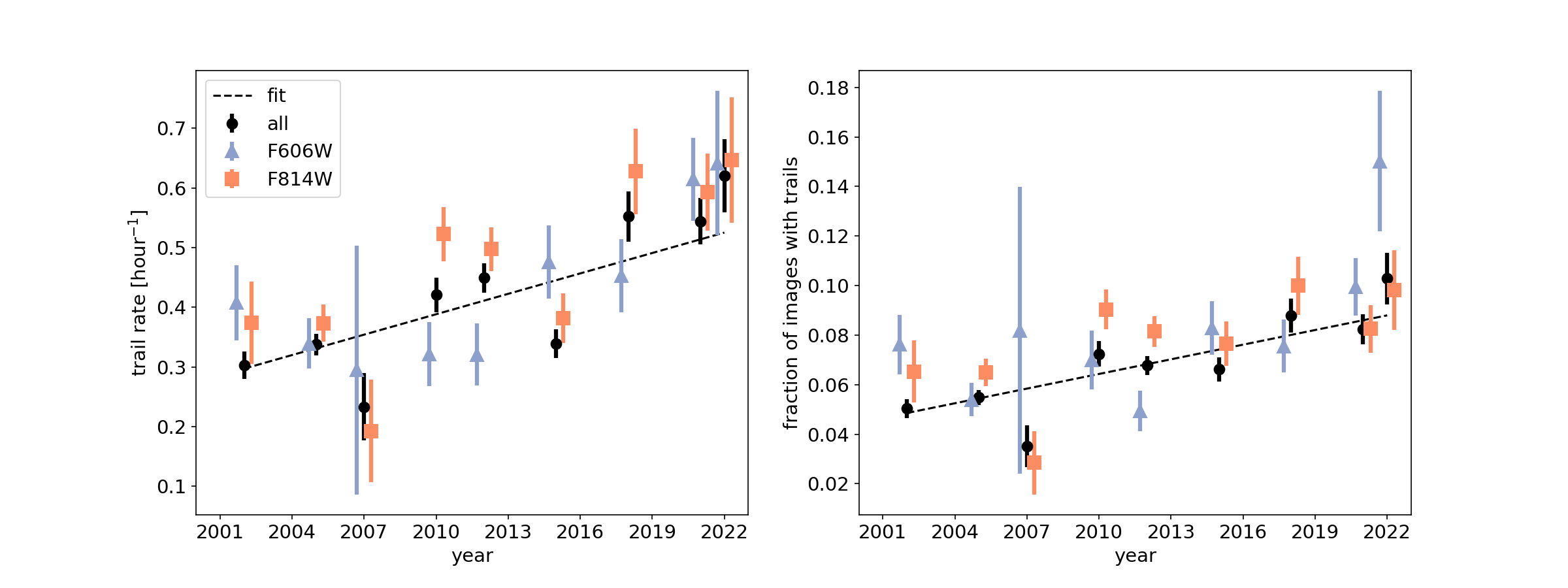}
\caption{(left) Rate of satellite trails (in trails per hour) in ACS/WFC data as a function of year. Error bars come from Poisson statistics on the number of trails. (right) Fraction of ACS/WFC images with satellite trails as a function of year. Both panels also plot the data for just the F606W and F814W filters, which show similar trends as the full data set. Linear fits to the data from all filters are provided by the black-dashed line. Both the trail rate and fraction of images affected has increased by a factor of $\sim2$ since the start of ACS observations.}
\label{fig:trails_vs_time}
\end{figure}

Table~\ref{tbl:trail_rate} provides the number of ACS/WFC images searched and the number of trails found as a function of year. Even though we analyze each WFC chip independently, any trail that crosses both chips (and is therefore detected twice) is only counted as one trail. In addition to providing the fraction of images affected by satellite trails, we calculate the satellite trail incidence rate as
\begin{equation}
\rm{trail\,rate}=\left(\frac{\rm total\,number\,of\,trails}{\rm total\,exposure\,time}\right)\eta_{\theta}\eta_{L}
\end{equation}
where $\eta_{\theta}$ and $\eta_L$ are multiplicative correction factors to account for the range of angles and trail lengths not considered when summing the number of trails. These correction factors assume the true location and orientation of satellite trails is random. In our case, $\eta_{\theta}=1.08$ and $\eta_L=1.14$. Results are also shown in Figure~\ref{fig:trails_vs_time}. 
 
We find a general increase in both the fraction of images affected by satellite trails and the incidence rate by approximately a factor of two between 2002 and 2022. While there are some large jumps from year to year (e.g., the significant drop in trails in 2007 is related to the lack of available data due to the electronics failure during that year), the general trend is towards more data being contaminated over time.

If the detection of satellites were a function of wavelength, the evolution in satellite incidence rate could reflect the evolution in the most frequently used filters over time. To account for this possibility, we also plot the satellite incidence rates considering only images taken with the F606W and F814W filters. These filters were chosen because they are consistently the most heavily used over the lifetime of ACS, so provide large samples even though we are downsizing our data set. The two filters still show the same general trend as the complete data set.  

\begin{table}
\centering
\begin{threeparttable}
\caption{Summary of satellite trails identified in ACS/WFC data.}
\begin{tabular}{ || c c c c c c || }
\hline
year & Images & $\sum{\rm EXPTIME}$ & Images w/ Trails & Trails& Trail Rate $^\dagger$  \\
     &        & (hours)             &                    &                   & (hours$^{-1}$) \\
\hline
2002 & 4053 & 704.94 & 166 & 174 & 0.30$\pm$0.02\\
2005 & 7479 & 1241.45 & 334 & 341 & 0.34$\pm$0.02\\
2007 & 594 & 89.62 & 17 & 17 & 0.23$\pm$0.06\\
2010 & 3358 & 610.61 & 198 & 209 & 0.42$\pm$0.03\\
2012 & 5729 & 915.97 & 316 & 335 & 0.45$\pm$0.02\\
2015 & 3341 & 716.93 & 180 & 198 & 0.34$\pm$0.02\\
2018 & 2306 & 389.59 & 165 & 175 & 0.55$\pm$0.04\\
2021 & 2714 & 444.87 & 182 & 197 & 0.54$\pm$0.04\\
2022 & 1159 & 201.94 & 97 & 102 & 0.62$\pm$0.06\\
\hline
\end{tabular}
\begin{tablenotes}
\footnotesize
\item $^\dagger$ Includes correction factors $\eta_{\theta}$ and $\eta_L$
 \end{tablenotes}
 \label{tbl:trail_rate}
\end{threeparttable}
\end{table}

\begin{figure}
\includegraphics[width=\columnwidth,center]{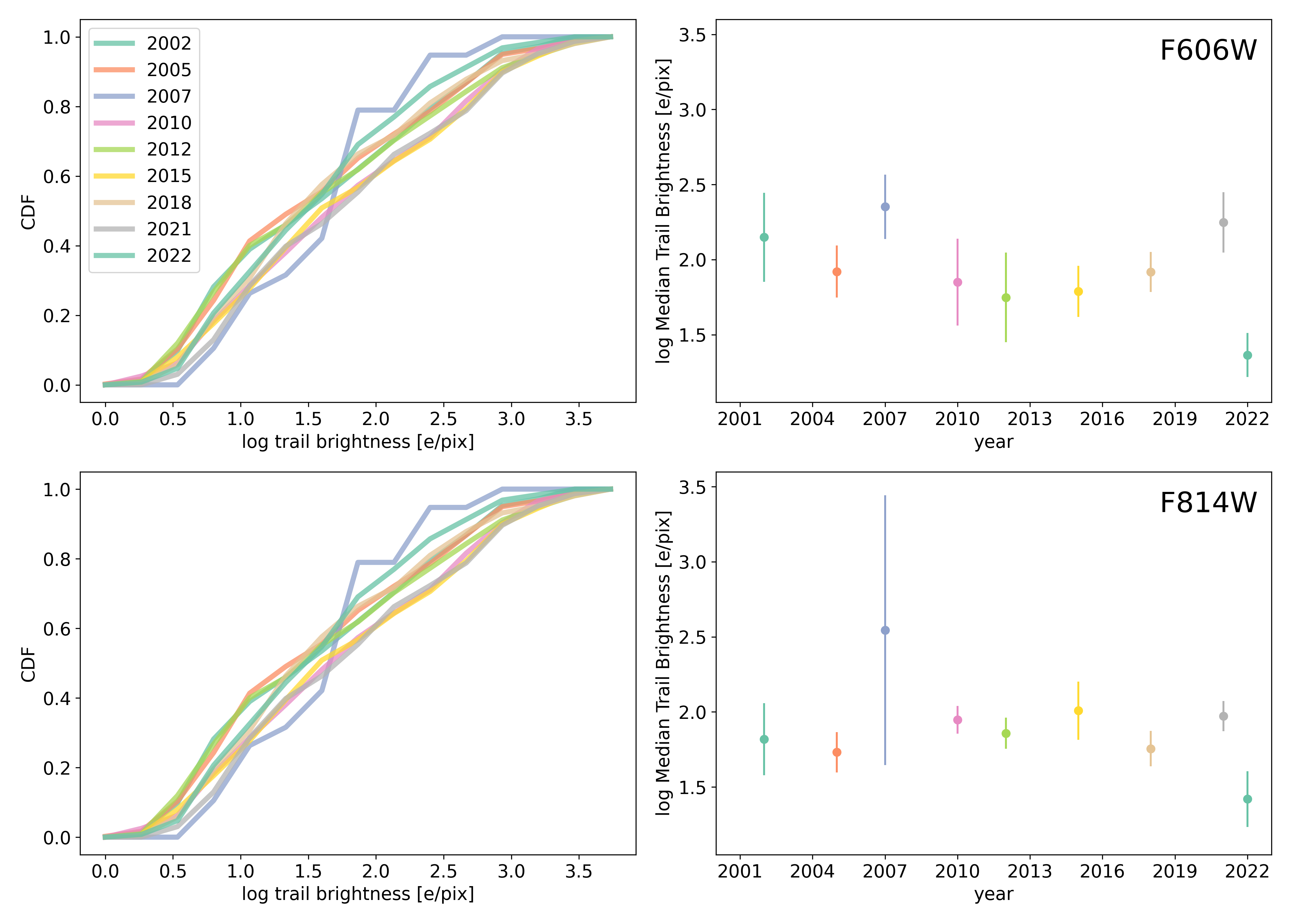}
\caption{(left) CDFs of trail brightnesses measured for each year. (right) The median trail brightness as a function of time. There is no clear systematic variation in the brightness of satellites over time.}
\label{fig:brightness_vs_time}
\end{figure}

We also investigate how the brightness of satellite trails has evolved. We measure the mean trail brightness as the average intensity measured over the length of the trail and within $\pm$ 1-$\sigma$ of the Gaussian fit to the cross-section of each trail. The results are in Figure~\ref{fig:brightness_vs_time}, where both the trail brightness cumulative distribution function (CDF) and median brightness as a function of time are plotted. Once again, any evolution in satellite trail brightness over time could be affected by any change in the filters being used\footnote{Indeed, when measuring the mean brightness including all filters together, there is a jump in mean brightness after 2007.}. We therefore analyze only the F606W and F814W filters. At fixed filter, we see no systematic trend in the brightness distributions of satellites, although some brightness distributions are distinct at a statistically significant level ($P<0.003$) according to K-S tests. 

\section{Discussion and Conclusions} \label{s:conclusion}
In this report, we present a new method of identifying and masking satellite trails and other linear artifacts in ACS/WFC imaging data using the Median Radon Transform (MRT), a modified form of the standard Radon Transform. The key advantages of this approach are its high sensitivity and robustness against bright astronomical sources in images, as long as an image is sufficiently dominated by the background. We have shown how this approach can be sensitive to trails significantly below the background noise level of images. Tests on real data show it recovers $\sim$85\% of trails identified by-eye with a false detection rate of 2.5\%.

We also use a large collection of images taken over two decades two infer which regions of parameter space are most prone to false detections. We showed how dense fields and image corners (where linear paths across an image are shortest) are most prone to false trails caused by the chance alignment of astronomical sources. We have also demonstrated that our approach cannot explicitly differentiate between satellite trails and other artifacts such as diffraction spikes and scattered light, but these may be distinguishable by their orientations and/or presence in multiple images of the same field. We estimate that if one avoids parameter space where false trails dominate, the false detection rate is $\sim10\%$. 

The Python package developed for this study (\texttt{findsat{\_}mrt}) is now available as part of the \texttt{acstools} package. The methodology presented here is a first attempt at using the MRT to identify satellite trails in imaging data, and a number of improvements can be made. From a practical standpoint, the MRT is expensive to calculate. Methods to dramatically decrease calculation time exist \citep{Press2006, Nir2018} and could be applied here. Our algorithm currently uses a number of tuned parameters to filter-out false detections after they are extracted from the MRT. Further exploration into simpler approaches to discarding these false detections, perhaps using a direct analysis of their shapes in the MRT itself (analogous to how programs like \texttt{SExtractor} differentiate extended and point sources; \citealt{Bertin1996}), could be worth undertaking. On a general note, explicitly better ways to reject false trails caused by the alignment of astronomical sources will broaden the applicability of our approach. Lastly, in the future we plan to rerun this algorithm on distortion-corrected individual exposures, which will increase sensitivity and remove the need to match trails that cross multiple chips in the same image. 

Thankfully, we have been able to identify regions of parameter space that yield relatively clean samples of satellite trails, and we have analyzed how the rate and brightness of satellites has evolved in the 20 years of ACS operations. We find the rate of satellite trail contamination has increased by approximately a factor of two over this time span, while the brightness of satellite trails shows no clear evolution.

While the rate of satellite trails affecting ACS/WFC data has increased, does not appear to have risen as dramatically as the number of human-made satellites in orbit over the last few years\footnote{\url{https://www.ucsusa.org/resources/satellite-database}}. This difference may have to do with the orbital properties of HST and other satellites, but we defer a thorough analysis of expected contamination rates due to the artificial satellite population to future work. Regardless, as the rate of artificial satellites in orbit continues to grow, reliable satellite trail identification and masking will be even more important.

\section*{Acknowledgements}
We thank Meaghan McDonald, Roberto Avila, Yotam Cohen, and Gagandeep Anand for helpful feedback on this report. We also thank Anton Koekemoer for his assistance retrieving the by-eye satellite masks for the Frontier Fields program. This study made use of the \href{https://scikit-image.org/}{scikit-image} \citep{scikit-image}, \href{https://numpy.org/}{Numpy} \citep{numpy}, \href{https://www.astropy.org/index.html}{Astropy}  \citep{astropy:2013,astropy:2018,astropy:2022}, \href{https://matplotlib.org/stable/index.html}{Matplotlib} \citep{matplotlib}, \href{https://github.com/astropy/photutils}{photutils} \citep{photutils}, 
 and \href{https://github.com/astropy/ccdproc}{ccdproc} \citep{ccdproc} Python packages.

\bibliography{miles_cmdtime}
\bibliographystyle{apj}

\end{document}